\documentclass{iopart}

\usepackage{amssymb}
\usepackage{graphicx}

\newcommand{\dii}{{\rm d}}
\newcommand{\ii}{{\rm i}}
\newcommand{\bsq}{B,S,Q}

\begin{document}

\title{The canonical effect in statistical models for
relativistic heavy ion collisions}

\author{A Ker\"anen\dag and F Becattini\ddag}

\address{\dag\ Department of Physical Sciences,\\
P.O. Box 3000, FIN-90014 University of Oulu, Finland}
\address{\ddag\ Universit\`a di Firenze and INFN Sezione di Firenze,\\
Via G. Sansone 1, I-50019 Sesto F.no, Firenze, Italy}

\ead{\dag\ antti.keranen@oulu.fi,
\ddag\ becattini@fi.infn.it}

\begin{abstract}
Enforcing exact conservation laws instead of average ones 
in statistical thermal models for relativistic heavy ion reactions
gives raise to so called {\em canonical effect}, 
which can be used to explain some enhancement effects
when going from elementary (e.g. pp) or small (pA) systems
towards large AA systems.
We review the recently developed method for computation
of canonical statistical thermodynamics, and give an insight when
this is needed in analysis of experimental data.  
\end{abstract}

\submitto{\JPG}

\pacs{24.10.Pa, 25.75.-q, 25.75.Dw}

\maketitle

\section{Introduction}

Statistical thermal models are widely used in the analysis
of relativistic heavy ion collisions (see e.g. 
\cite{features, cleymans319, heppe, cleymans5284, cleymans3319}, and
references therein).
In these models, the final stage of inelastic interactions
between hadrons, the {\em chemical freeze-out},
is characterized by the parameters of stationary
maximum entropy ensemble. 

For large system, it is appropriate to choose the most 
straightforwardly applicable ensemble, the grand canonical (GC) one,
which is parametrized by temperature $T$ and chemical potential
$\mu_i$ for each averagely conserved charge $i$.
The GC ensemble is defined in the large volume limit,
so the volume parameter $V$ is an extensive coefficient
in the expressions for such quantities as mean particle numbers.
For small systems, such as pp or pA reactions, the charge conservation
laws must be handled exactly, so the ensemble to be chosen is
the canonical (C) one. Then, the $V$ independent 
fugacities due to average conservation laws do not appear,
but are replaced by {\em nonlinearly $V$ dependent} canonical 
chemical factors, which can account for many enhancement patterns
going from pp to pA to AA systems \cite{cleymans2747, redlich413, kebe}.

In this paper, we quote the newly developed computational
method for relativistic $\bsq$ canonical hadron ensemble, which enables
us to reach previously unattainable canonical model results,
up to baryon number ${\cal O}(100)$ whereas with old methods it was
 only possible to reach
$B\sim 20$ \cite{cleymans2747} with very long computation times.
Once this is achieved, we can give an answer to question
whether the full $\bsq$ canonical modelling is needed,
or is the $S$ canonical ($B,Q$ grand canonical)
or GC approximation accurate enough.

\section{Methods}

In order to pass from the usual grand-canonical partition function 
$Z_{GC}$ to the 
one fulfilling exact internal symmetries, we employ a well-known group
theoretical 
method, first introduced by Cerulus \cite{cerulus, turko201, turko153}.
By denoting the set of conserved quantum numbers by $\{C_i\}$, the canonical
partition function $Z_{\{C_i\}}$ can be obtained by using a projecting 
operator 
onto the conserved quantum numbers. In the case of $N$ internal symmetries of
type
$U(1)$, the projection takes the form:
\begin{equation}
Z_{\{C_i\}}(T,V) = \left[ \prod_{i=1}^{N} 
\frac{1}{2\pi}  \int_0^{2\pi}\dii\phi_i e^{-\ii C_i\phi_i}\right]
         Z_{GC}(T,V,\{\lambda_{C_i}\}).
\label{eq:projection}
\end{equation}
where $\phi_i \in [0,2\pi)$ is a $U(1)$ group parameter and a Wick-rotated 
fugacity factor $\lambda_{C_i} = e^{i\phi_i}$ is introduced, for every charge 
$C_i$. 

In heavy ion reactions, the relevant set of conserved charges is $\{C_i\} = B,
S, Q$, namely baryon number, strangeness and electric charge.
Applying this set to the previous equation yields a triple integral,
which is computationally extremely time consuming for $B \gtrsim 5$.
In \cite{kebe} we have eliminated analytically the baryon integration,
and have obtained a very efficient way to calculate 
the full canonical ($\bsq$) hadron thermodynamics.
The applied form of partition function reads:
\begin{eqnarray}
Z_{B,S,Q}(T,V) &=& \frac{Z_0}{(2\pi)^2} \int_0^{2\pi}\dii\phi_S 
\int_0^{2\pi}\dii\phi_Q \\
&\times&\cos\left(S\phi_S+Q\phi_Q+B\arg \omega(\phi_S,\phi_Q) \right)
\nonumber \\ 
&\times&I_B(2|\omega(\phi_S,\phi_Q)|) \nonumber \\
&\times& \exp\left[2\sum_M z_i^1\cos(S_i\phi_S +
Q_i\phi_Q)\right], \label{partt}\nonumber
\end{eqnarray}
where $Z_0$ denotes the partition function for hadrons carrying no
relevant charge and $I$ is the modified Bessel function.
The sum is over mesons, and $z^1_i$ is the one-particle partition
function.
The $\omega$ above is defined as 
$\sum_B z_i^1 \, e^{\ii(S_i\phi_S+Q_i\phi_Q)}$, where the summation
runs over baryons but not antibaryons.
Now, we are left with a double integration only which can be then performed 
numerically with no major problem.

The mean particle numbers of primary hadrons (i.e. those directly emitted from
the hadronizing source) $\langle N_i \rangle$ are obtained from the 
equation~(\ref{partt}) \cite{hagedorn541,becattini269} by taking the 
derivative 
of the canonical partition 
function with respect to a fictitious fugacity $\lambda_i$:
\begin{eqnarray} 
\langle N_i \rangle &=& \left. \lambda_i \frac{\partial
\ln Z_{B,S,Q}(T,V)} {\partial \lambda_i} \right|_{\lambda_i =1} 
\nonumber \\
&=& \frac{Z_{B-B_i,S-S_i,Q-Q_i}(T,V)}{Z_{B,S,Q}(T,V)} z_i^1,
\label{Cmean}
\end{eqnarray}
where the quantity $Z_{B-B_i,S-S_i,Q-Q_i}/Z_{B,S,Q}$ is called {\em chemical
factor} 
\cite{becattini269}. In the large volume (thermodynamical) limit, this
expression
becomes the GC one, i.e. 
$\langle N_i \rangle = \lambda_B^{B_i}\lambda_S^{S_i}\lambda_Q^{Q_i}z_i^1$.

\section{Numerical results}

In this paper, we perform no actual fitting procedure 
to the experimental data, but only use some suggestive
thermal parameter values from our previous work \cite{features}.
This is because our main motivation is to show systematically
the canonical effects in different systems in order to
give the reader an insight on applicability of different
thermal approaches, namely full canonical, strangeness canonical
and grand canonical.

In the following, we fix the temperature and baryon density in
several cases. Strangeness and charge to baryons ratio are fixed
according to initial nuclear composition.
These conditions lead to fixed fugacities and particle ratios
in GC formalism. GC approximations are then to be compared
with full canonical and $S$-canonical counterparts, 
which are nonlinearly dependent on net baryon content. 

\begin{figure}
\begin{center}
\includegraphics[scale=0.28, angle=-90]{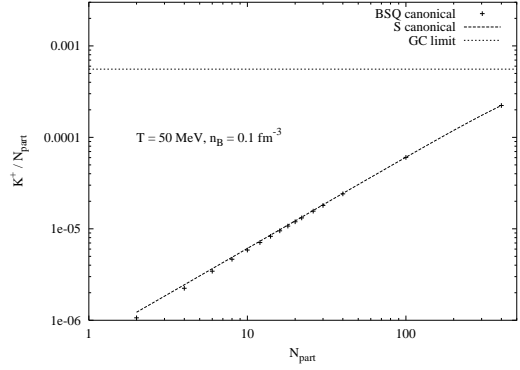}
\end{center}
\caption{Canonical enhancement of kaons as a function of number of 
participants 
in the conditions relevant to GSI energies.}
\label{fig1}
\end{figure}

In figure \ref{fig1} we show the canonical kaon enhancement
in the conditions relevant to GSI energies, 1.7 GeV per nucleon
in SIS AuAu collisions.
The theoretical production ratio K$^+$ over number of participating 
nucleons is shown to point out the canonical strangeness effect. Note that we
have been able to compute the full chemical factor even for $B=400$.
The ratio increases very slowly towards the GC limit, thus strangeness must
always be 
handled canonically when analysing SIS results. The relative 
difference between 
full canonical and strangeness-canonical results is 13\% for $B=2$, decreasing
to 4\% for
$B=10$ and to 1\% for $B=40$. It must be pointed out that the 
above results are calculated using an isospin symmetric initial 
configuration whereas $Z/A\simeq 0.4$ for gold nucleus. 
However, this variation essentially gives no change on the
relative differences above, although the ratio K$^+/N_{\mbox{{\tiny part}}}$ 
decreases naturally due to the initial neutron excess.  

\begin{figure}
\begin{center}
\includegraphics[scale=0.28, angle=-90]{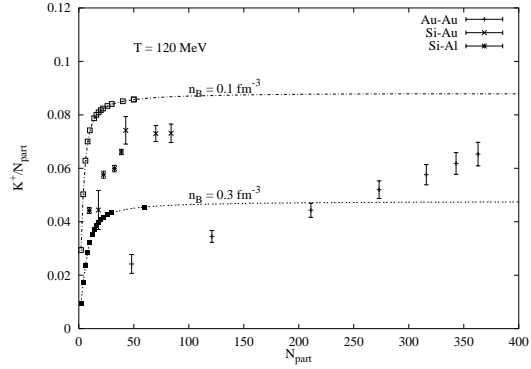}
\end{center}
\caption{Theoretical K$^+/N_{\mbox{{\tiny part}}}$ curves at fixed temperature 
for two different baryon densities shown along with AGS experimental results 
\cite{ahle}. Curves are strangeness canonical while squares are full canonical
results}
\label{fig2}
\end{figure}

In Au--Au collisions at AGS energies (11.6 $A$ GeV momentum), the chemical 
freeze-out temperature is found to be around $T=120$ MeV and the 
baryon density  close to the normal nuclear density \cite{features}.
In figure \ref{fig2} we plot the 
calculated K$^+/N_{\mbox{{\tiny part}}}$ with 
results obtained in peripheral to central 
Au--Au, Si--Au and Si--Al collisions.
It can be seen that all the way up from $B=2$ to $B=60$ the full canonical and
strangeness-canonical results are essentially the same. All central reaction 
results lie on the region where the canonical effects are negligible 
and the GC formalism applies.
Strangeness enhancement in Au--Au system does not definitely look like being
of canonical origin, whereas Si--Au and Si--Al follow roughly the curve at the 
normal nuclear density (not shown in the figure). The above argument also
applies for the K$^-/N_{\mbox{{\tiny part}}}$ enhancement pattern
\cite{kebe}.

\begin{figure}
\begin{center}
\includegraphics[scale=0.28, angle=-90]{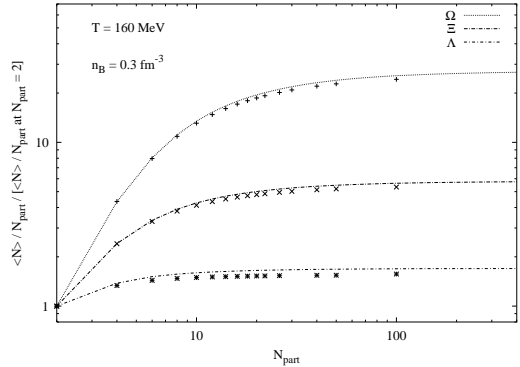}
\end{center}
\caption{Strange baryon enhancement in the conditions relevant to the SPS
energies. 
Hadron multiplicities are normalized to
results at $B=2$. Curves are strangeness 
canonical while crosses are full canonical results.} 
\label{fig6}
\end{figure}

The best fit temperature for the multiplicities measured by NA49 
experiment at SPS 
in central Pb--Pb reactions at 158 $A$ GeV is found to be $T\sim160$ MeV
\cite{features}.
In figure \ref{fig6} we show the strange baryon enhancement by using a baryon
density 0.3 fm$^{-3}$. 
Although the baryon density in our analysis \cite{features} was found
to be $n_B \sim 0.2$ fm$^{-3}$ this larger value is used in order to probe the 
largest reasonable canonical effect. 
All results in figure \ref{fig6} are normalized to the  
    baryon multiplicities per participant at the point
$N_{\mbox{{\tiny part}}} = 2$. This choice reveals the slight
    difference between the results obtained using the $S$-canonical
    approximation and the $\bsq$-canonical calculation.
An interesting feature here is the fact, that the $S$-canonical
approximation leads to an overestimation of the canonical effect.



\ack
We would like to acknowledge stimulating discussions
with Esko Suhonen, Jean Cleymans, Krzysztof Redlich and many other 
colleagues participating the conference Strangeness in Quark Matter
2001, and especially the organizers of that pleasant meeting. 
One of us (A.K.) 
acknowledges the support of the Bergen Computational
Physics Laboratory in the framework of the European Community - Access
to Research Infrastructure action of the
Improving Human Potential Programme. 



\end{document}